# Local Structure of the Superconductor $K_{0.8}Fe_{1.6+x}Se_2$: Evidence of Large Structural Disorder

T. A. Tyson[1], T. Yu[1], S. J. Han[2], M. Croft[3], G. D. Gu[2], I. K. Dimitrov[2], and Q. Li[2]

[1]Department of Physics, New Jersey Institute of Technology, Newark, NJ 07102, USA

[2]Condensed Matter Physics and Materials Science Department, Brookhaven National Laboratory, Upton NY 11973, USA

[3]Department of Physics and Astronomy, Rutgers University, Piscataway, NJ 08854, USA

## Abstract

The local structure of superconducting single crystals of $K_{0.8}Fe_{1.6+x}Se_2$ with Tc = 32.6 K was studied by x-ray absorption spectroscopy. Near-edge spectra reveal that the average valence of Fe is 2+. The room temperature structure about the Fe, K and Se sites was examined by iron, selenium and potassium K-edge measurements. The structure about the Se and Fe sites shows a high degree of order in the nearest neighbor Fe-Se bonds. On the other hand, the combined Se and K local structure measurements reveal a very high level of structural disorder in the K layers. Temperature dependent measurements at the Fe sites show that the Fe-Se atomic correlation follows that of the Fe-As correlation in the superconductor $LaFeAsO_{0.89}F_{0.11}$ - having the same effective Einstein temperature (stiffness). In $K_{0.8}Fe_{1.6+x}Se_2$, the nearest neighbor Fe-Fe bonds has a lower Einstein temperature and higher structural disorder than in $LaFeAsO_{0.89}F_{0.11}$. The moderate Fe site and high K site structural disorder is consistent with the high normal state resistivity seen in this class of materials. For higher shells, an enhancement of the second nearest neighbor Fe-Fe interaction is found just below Tc and suggests that correlations between Fe magnetic ion pairs beyond the first neighbor are important in models of magnetic order and superconductivity in these materials.

# I. Introduction

The observation of superconductivity in the quaternary ZrCuSiAs-type systems (1111 type) iron arsenide (pnictide) system $LaFeAsO_{1-x}F_x$ [1] created a renaissance in research in superconductivity from both the applied and fundamental physics perspectives. Over the last three years extensive studies have been conducted on the Fe based systems and the results have been reviewed in recent articles (See Ref. [2]). In the $RFeAsO_{1-x}F_x$ (R=rare earth) 1111- type system, optimization of the chemical properties led to the realization of a superconducting transition temperature of ~55 K in $SmFeAsO_{1-\delta}$ [3] , the highest in these new iron systems to date. This first class of materials possesses normal state resistivity values near the transition temperature which are less than 1 mΩ.cm with linear behavior at higher temperatures. Following this, superconductivity was discovered in the system $AFe_2As_2$ system (A=K, Se, Ba, called the 122 system) [4] with an ambient pressure transition temperature, Tc ~38 K, and resistivity near the transition temperature is ~1 mΩ.cm. Another class of materials referred to as the 111 type (with $CuSb_2$ structure), LiFeAs, was observed to superconduct with a transition near ~18 K [5]. More recently, the PbO type system such as $FeSe_{1-x}$ (with defects on the Fe and Se sites) as well as the Te alloys of this system were found to exhibit superconductivity with Tc near 8 K at ambient temperature for $FeSe_{1-x}$ [6]. Tc was found to be optimized near ~37 K for an external pressure of ~7 GPa [7]. In this system the resistivity values just above the onset of superconductivity are also ~1 mΩ.cm and linear behavior is exhibited above Tc.

Most recently, superconductivity was observed in the system $A_XFe_ySe_2$ (with Tc ~ 31 K in $K_{0.8}Fe_{1.6+x}Se_2$) [8] and has enhanced interest in the field by virtue of the fact the Fe sites possess high moments and hence the possibility of coexisting antiferromagnetic state and superconducting state is raised. This class of $A_XFe_ySe_2$ (A = alkali or Tl) materials differs from the previous systems in many significant ways. The resistivity just above the transition is > 10 mΩ.cm (more than 10 times higher than that of other iron based systems) and in addition it displays a maximum in the normal state resistivity vs.

temperature curve in the region between ~100 K and 300 K. The magnetic moment on the Fe site is ~3.31 $\mu_B$, the largest of the FeAs and FeSe based systems, and antiferromagnetic order onsets near ~ 550 K. There is evidence for ordered Fe vacancies [9]. $^{57}$Fe Mossbauer spectroscopy measurements indicate that the ordered magnetic state persists below Tc [10]. X-ray diffraction measurements on single crystals suggest an intrinsic phase separation between a majority non-superconducting $\sqrt{5}$ x $\sqrt{5}$ x 1 Fe defect ordered phase and a minority $\sqrt{2}$ x $\sqrt{2}$ x 1 superconducting phase [11].

Understanding the structural changes at the distinct Fe, Se and K ion sites is central to distinguish the important structural components which support superconductivity, when compared with the better characterized $LaFeAsO_{1-x}F_{x.}$ system. Hence, the local structure of superconducting single crystals of $K_{0.8}Fe_{1.6+x}Se_2$ with Tc = 32.6 K was studied by x-ray absorption spectroscopy. Near-edge spectra reveal that the average valence of Fe is 2+. The room temperature structure about the Fe, K and Se sites was examined by iron, selenium and potassium K-edge measurements. The structure about the Se and Fe sites shows a high degree of order in the nearest neighbor Fe-Se bonds. For higher shells, enhancement of the second nearest neighbor Fe-Fe interaction is found just below Tc and suggests that correlations between Fe magnetic ion pairs beyond the first neighbor are important.

## II. Experimental Methods

High quality single crystal samples of $K_{0.8}Fe_2Se_2$ synthesized by the unidirectional solidification method [12], were extracted 60 to 70 mm from the edge of a 200 mm long crystal bar and were characterized by magnetization and magneto-resistivity measurements. The onset of the transition was at $T_c$ = 32.6 K, with a transition width $\Delta T_c$ = 0.3 K (10%-90%) , as seen in Fig. 1(b). X-ray absorption spectra were measured in florescence mode at the National Synchrotron Light Source beamlines X3B (Fe K-edge, 19 K to 300 K), X11A (Se K-edge, room temperature) and X15B (K K-ege, room temperature) at Brookhaven National Laboratory. To reduce the possibility of reaction of the samples with oxygen or moisture, samples were kept in pure Ar (99.9999%) environment in all time prior to quick , transfer of the samples from the glove box to the experimental x-ray sample chambers. Fe K-Edge measurements were conducted with the sample under vacuum conditions with base pressures < $10^{-6}$ millibar. Measurements at the K K-edge were conducted in a He environment and the Se K-edge measurement was conducted with the sample in a vacuum sealed container. No changes in x-ray spectra were found between successive data scans. Also, no changes were found when comparing data taken at the beginning and end of the complete measurements cycles.

The Se K-edge measurements were conducted with a Lytle type fluorescence detector using an As Z-1 filter (6 absorption lengths) for elastic scatter suppression. The K K-edge spectra were collected with a Si(Li) single element solid state detector and the Fe K-edge data were collected with at 31 Element Ge solid detector using Mn Z-1 Mn filters (9 absorption lengths) for elastic scatter suppression. All data were corrected for self-absorption using the method of Ref. [13]. The measurements were conducted with the single crystal c-axis held ~45° to the incident x-ray beam with the crystal c-axis in the plane of the synchrotron ring. Temperature dependent Fe K-edge measurements were made on warming the single crystal from 19 K on the cold finger of a Displex™ cryostat. The uncertainty in temperature is < 0.25 K. Two to six scans were taken at each temperature. A Fe foil reference was employed for energy calibration at the Fe K-Edge. The reduction of the x-ray absorption fine-structure (XAFS) data was performed using standard procedures [14]. For the fits to the Fe K-edge temperature dependent data, to treat the distribution on equal footing at all temperatures, the spectra were modeled in R-space by optimizing the integral of the product of the radial distribution functions and theoretical spectra with respect to the measured spectra [15] at each temperature as done in Ref [16]. Theoretical spectra for atomic shells [17] were derived from the crystal structure data [18]. The predicted trends in the fits and model spectra of Fig. 3 were found to be independent of use of the long range order based on the space groups I4/mmm or I4/m with Fe ordering. For the Fe K-Edge r-space fits of the Fe-Se and Fe-Fe distribution were confined to the k-range $2.54 < k < 12.8$ Å$^{-1}$ and to R-range $1.48 < R < 4.60$ Å (with $S_0^2 = 0.73$).

In the temperature dependent fits at Fe K-edge, the coordination numbers were fixed while varying the width and positions of the Gaussian components of the radial distribution functions. Errors reported are due to the statistical errors based on the spread of parameters over the consecutive scans at fixed temperature. The temperature dependence of the Fe-Se and Fe-Fe Debye-Waller factors ($\sigma^2$) were modeled by a static contribution ($\sigma_0^2$) plus a single parameter ($\theta_E$) Einstein model [19]. In Fig. 3, model curves computed for the Fe, Se and K sites with the ordered $K_{0.8}Fe_{1.6+x}Se_2$ system using $S_0^2 = 0.8$ and a global $\sigma^2$ values of 0.006 Å$^2$, which are reasonable for room temperature estimates. In this figure, for the Se K-Edge and the K K-edge the Fourier transforms of the model and data were over the k- ranges $2.63 < k < 17.14$ Å$^{-1}$ and $1.98 < k < 10.0$ Å$^{-1}$ , respectively. Note that the peaks in the Fourier transforms, Figs. 3, 4 and 5(b), are at shorter distances than the corresponding bond distances do to the central atom phase shifts and the scattering atom phase functions. Accurate distances are obtained by model fits.

## III. Results and Discussion

To determine the valence of the sample the near edge spectra or threshold spectra (called x-ray absorption near edge spectrum XANES) were measured with as step size of 0.2 eV to bring out features in the main line 1s $\rightarrow$ "4p" peak. The main line spectra are presented as the thick line in Fig. 2(a) and are compared with a group of ~4+, 3+ and 2+ Fe compounds standards. The main-edge energy shows a chemical shift to lower energy with decreasing valence. The chemical shift of the $K_{0.8}Fe_{1.6+x}Se_2$ spectrum falls clearly in the group of Fe 2+ standards. The lack of sharp features of the $K_{0.8}Fe_{1.6+x}Se_2$ spectrum is consistent with broad bands in the Fe-site p-symmetry projected DOS.

The Fe-K pre-edge region, below 7.12 keV, is dominated by 1s transitions into final d-states with the 1s-hole/3d-electron final state Coulomb interaction being what shifts these transitions below the main edge. In Fig. 2(b) the pre-edge spectrum for the same set of samples from the previous figure are shown. One can see a systematic chemical shift of the pre-edge feature from the "a 2+" to the "b 3+" and finally to the "c ~4+" energy range with increasing Fe valence in the compounds. The $K_{0.8}Fe_{1.6+x}Se_2$ pre-edge clearly falls in the "a 2+" energy range. In general for a 3d transition metals in centrosymmetric local environment the quadrupole allowed 1s to 3d pre-edge transitions increase in intensity with increasing valence. Note that d-p hybridization can, however, enhance the pre-edge feature intensity by introducing stronger dipole allowed transitions. The tetrahedral Fe-Se environment in $K_{0.8}Fe_{1.6+x}Se_2$, and the Fe-S environment in Fe-S-en (en= ethylenediamine) [20], are non-centrosymmetric with d-p hybridization allowed and their pre-edges are both seen to be substantially enhanced in intensity.

In Fig. 1(a) we show the crystal structure of $K_{0.8}Fe_{1.6+x}Se_2$ without defects on the Fe and K sites for reference to the structural discussions. The local structure about the Fe, Se and K sites was examined by x-ray room temperature at the iron K-edge, the selenium K-edge and the potassium K-edge measurements. In Fig. 3 we show the curves of the measured data and a corresponding model based on diffraction models as mentioned above. The experimental data are displayed as solid lines and the model curves are displayed as dashed lines. With respect the structure about Fe (Fig. 3(a)), the first peak in the Fourier transform (XAFS structure function) has two components from Fe-Se and Fe-Fe (first Fe-Fe correlation) bonds. While the model curve exhibits a shoulder on high R side of the first peak, corresponding to Fe-Fe, this peak is suppressed in the experimental spectrum. As will be seen in the temperature dependent data below, this is due to static disorder in the Fe layer. The Fe-Se nearest neighbor peak has a profile which matches qualitatively that of the model, on the other hand. Beyond the first peak there are higher order shells corresponding to the second neighbor Fe-Fe bond, the Fe-K bond and the Fe-Se bond. Analysis of consecutive scans and adjustments of the Fourier transform range to ascertain truncation effects reveal that the weak peak as the second neighbor Fe-Fe bond, the Fe-K bond and the Fe-Se bonds in the data are real. However, they are suppressed indicating a high level of disorder in this material. Note that all of these bonds correspond to the same $Fe_2Se_2$ layer.

With respect to the Se sites one can see in Fig. 3 that the first shell about Se is Fe and the second shell would contain K. In Fig. 3(b) we see the structure function for the local structure with respect to the Se sites. The first peak, composed only of Se-Fe bonds, is a close match with the qualitative model and the data show that the Fe-Se bonds have a low level of disorder. The second shell about Se corresponds to the Se-K bond and in this region there is negligible amplitude. This indicates a very high level of disorder in the K layers which will be seen in the K K-edge measurements. Near the Se-Se shell some non negligible amplitude is present while none exists for the Se-Fe peak.

In Fig. 3(c) we see the local structure about the K site compared to the model. A very low signal for the structure about K sites is measured. The first neighbor (typically the dominant XAFS signal) is Se, as can be seen from Fig. 1(a). In addition, no signals for higher order peaks such as K-Fe are found. The high level of order of the Se site seen in the Fe K-edge measurements and the absence of the Se-K peak combined with these results at the K K-edge support a model of very high structural disorder of the K sites (potassium layer). More information about the system with respect to the superconducting state and the static disorder in the Fe layer can be obtained from temperature dependent Fe K-edge x-ray measurements.

In Fig. 4, we show the Fourier transform data at the Fe K-edge between 27 K and 31.5 K. The region shown is between 3 Å and 3.7 Å (see Fig. 3(a) and the structural figure in 1(a)) and corresponds to the second neighbor Fe-Fe distance (see Fig. 2) in the same Fe layer. What is observed is that there is an enhancement of the Fe-Fe correlation near 29K, just below the transition to the superconducting state. The result suggests that understanding the magnetic order near the superconducting state requires a model which properly treats magnetic interaction between Fe sites significantly beyond the first neighbor Fe-Fe interactions .

We fit the first peak in the temperature dependent data between 19 K and 300 K to determine the behavior of the Fe-Se and Fe-Fe correlations for comparison with the $LaFeAsO_{0.89}F_{0.11}$ superconductor (Fig. 5). Typical consecutive scans at the Fe K-edge are shown in Fig 5(a) and a fit to the first peak (Fe-Se and Fe-Fe shells) in Fig. 5(b) for room temperature data. The temperature dependence of the Fe-Se and Fe-Fe Debye-Waller factors ($\sigma^2$) was modeled by an static disorder contribution ($\sigma_0^2$) plus a single parameter ($\theta_E$) Einstein model using the functional form $\sigma^2(T)=\sigma_0^2+\frac{\hbar^2}{2\mu k_B\theta_E}\coth(\frac{\theta_E}{2T})$ [8,21], where $\mu$ is the reduced mass for the bond pair. This simple model represents the bond vibrations as harmonic oscillations of a single effective frequency proportional to $\theta_E$. The parameter $\sigma_0^2$ represents the static disorder. It provides an approach to characterize the relative stiffness of the bonds and can be used to ascertain changes in pair correlations. It differs from the x-ray derived Debye-Waller factor in that the latter describes motion with respect to the equilibrium position of an atom.

The temperature dependence of the $\sigma^2$ for the Fe-Se bond is shown in Fig. 6 (a) and compared with that of the Fe-As bond in $LaFeAsO_{1-x}F_x$ from Ref. [16]. The temperature scale is a log scale to reveal the low temperature region. The $K_{0.8}Fe_{1.6+x}Se_2$ and the $LaFeAsO_{0.89}F_{0.11}$ systems have the same effective Einstein temperature for the first shell Fe-Se/As bonds. The static disorder ($\sigma_0^2$) parameter for the Fe-Se in $K_{0.8}Fe_{1.6+x}Se_2$ lies between that of the superconducting $LaFeAsO_{0.89}F_{0.11}$ system and the non-superconducting parent compound LaFeAsO. The result indicates that the is similar bonding in the Fe-As-Fe and Fe-Se-Fe networks in both systems.

With respect to the first neighbor Fe-Fe interactions there are some distinct differences between the $LaFeAsO_{1-x}F_x$ system and the superconducting $K_{0.8}Fe_{1.6+x}Se_2$ material. In Fig. 7 we see that while the first neighbor Fe-Fe interaction in $LaFeAsO_{1-x}F_x$ exhibits negligible (at the level of the data) static disorder, very significant disorder ($\sigma_0^2$ = 0.0069 Å$^2$) exists in the case of the $K_{0.8}Fe_{1.6+x}Se_2$. Moreover, the effective Einstein temperature is significantly lower for this bond than for the x=0.11 $LaFeAsO_{1-x}F_x$ system (226 K compared to 304 K). The lower Einstein temperature is a direct measurement of the softness of the Fe layer (showing that the Fe layer in $K_{0.8}Fe_{1.6+x}Se_2$ is softened compared to $LaFeAsO_{1-x}F_x$). Compared with the Fe-Se bond, the Fe-Fe first neighbor bond has asignificant temperature dependence over the measured range. These results compared with the enhancement of the second neighbor Fe-Fe interaction near Tc suggest that high order interactions J beyond the first neighbor must be considered to properly model the magnetism in these materials.

## IV. Summary

Near-edge spectra reveal that the average valence of Fe is 2+. The local structure about the Se and Fe sites shows a high degree of order in the nearest neighbor Fe-Se bonds. On the other hand, the combined Se and K local structure measurements reveal a very high level of structural disorder in the K layers. Temperature dependent measurements at the Fe sites show that the Fe-Se atomic correlation follows that of the Fe-As correlation in the superconductor $LaFeAsO_{0.89}F_{0.11}$ - having the same effective Einstein temperature (stiffness). In $K_{0.8}Fe_{1.6+x}Se_2$, the nearest neighbor Fe-Fe bonds has a lower Einstein temperature and higher structural disorder than in $LaFeAsO_{0.89}F_{0.11}$. The moderate Fe site and high K site structural disorder is consistent with the high normal state resistivity seen in this class of materials. For higher shells, an enhancement of the second nearest neighbor Fe-Fe interaction is found just below Tc and suggests that correlations between Fe magnetic ion pairs beyond the first neighbor are important in models of magnetic order.

## V. Acknowledgments

Support for this work was provided by the U.S. Department of Energy, Office of Basic Energy Science, Materials Sciences and Engineering Division under DOE-BES Grant DE-FG02-07ER46402 for T.A.T and T.Y. (NJIT) and under Contract No. DE-AC0298CH10886. DOE Grants (BNL), for S.J.H, G.G, I.K.D, and Q. L. Data acquisition was performed at Brookhaven National Laboratory's National Synchrotron Light Source (NSLS) which is funded by the U. S. Department of Energy.

## Figure Captions

**(Color online) Fig. 1**. (a) Crystal structure of defect free $KFe_2Se_2$ for the I/4m space group. For the superconducting materials it is suggested that random defects occur on the K sites and that the Fe1 sites are unoccupied – resulting in ordered Fe vacancies. (b) Resistivity curve for the single crystal samples.

**(Color online) Fig. 2**. Panel (a) shows the K-edge absorption spectrum of superconducting $K_{0.8}Fe_{1.6+x}Se_2$ compared to Fe systems of 2+, 3+ and 4+ valence states. In panel (b) the pre-edge region of the same spectra are also shown. The valence is seen to be strongly 2+ following the behavior of stoichiometric FeS.

**(Color online) Fig.3**. Local structure about the Fe, Se and K sites from the XAFS structure functions in panels (a), (b) and (c) respectively. In each panel, the data are represented by the solid curve and the dashed line a model with reasonable thermal/structural parameters. The components of the atomic shells are labeled in each panel. For the Fe local structure, the first neighbor Fe-Se and Fe-Fe peaks are well ordered while higher order correlations are weak. For the structure about Se, the Se-Fe bonds are well ordered while the Fe-K bonds are highly structurally disorders and the corresponding peak is unobservable. For the structure about K the structure peaks are significantly suppressed relative to the theoretical model for all distances. Only the first shell K-Se peak has non-negligible amplitude. This indicates that the K sites highly structurally disordered. Note that the peaks in the Fourier transforms, here and in Figs. 4 and 5(b), are at shorter distances than the corresponding bond distances due to the central atom phase shifts and the scattering atom phase functions. Accurate distances are obtained by model fits.

**(Color online) Fig. 4**. Temperature dependence of the second shell Fe-Fe peak shows enhancement of this second neighbor Fe-Fe shell in the Fe layer near the transition (~30 K). This suggests a connection between superconductivity and the magnetic ion interactions. Note that the peaks in the Fourier transforms are at shorter distances than the corresponding bond distances due to the central atom phase shift and the scattering atom phase functions. Accurate distances are obtained by model fits.

**(Color online) Fig. 5.** Two consecutive XAFS scans in k-space, at 300 K are given in (a) and the Fourier transform of the average data is shown with a fit to the Fe-Se and Fe-Fe (first neighbor) peaks in panel (b). The solid line corresponds to the data.

**(Color online) Fig. 6.** Extracted thermal parameters, $\sigma^2(T)$, for the (a) Fe-Se and (b)Fe-As first neighbor bonds in $K_{0.8}Fe_{1.6+x}Se_2$ and $LaFeAsO_{1-x}F_x$, respectively. The solid lines are with fits to Einstein models. Note the similarity between the two systems with respect to the first shell coordination of Fe.

**(Color online) Fig. 7.** Extracted thermal parameters, $\sigma^2(T)$, for the Fe-Fe second neighbor bonds in $K_{0.8}Fe_{1.6+x}Se_2$ (a) and $LaFeAsO_{1-x}F_x$ (b), respectively. Compared to the $LaFeAsO_{1-x}F_x$ system, $K_{0.8}Fe_{1.6+x}Se_2$ possesses significant static disorder in the Fe layer- consistent with the large normal state resistivity.

**(Color online) Fig. 8.** Extracted Fe-Se (solid circles) and Fe-Fe first neighbor (solid squares) bond distances showing the stronger temperature dependence (smaller Einstein temperature) for the Fe-Fe bond.

**Fig. 1. Tyson *et al.***

**(a)**

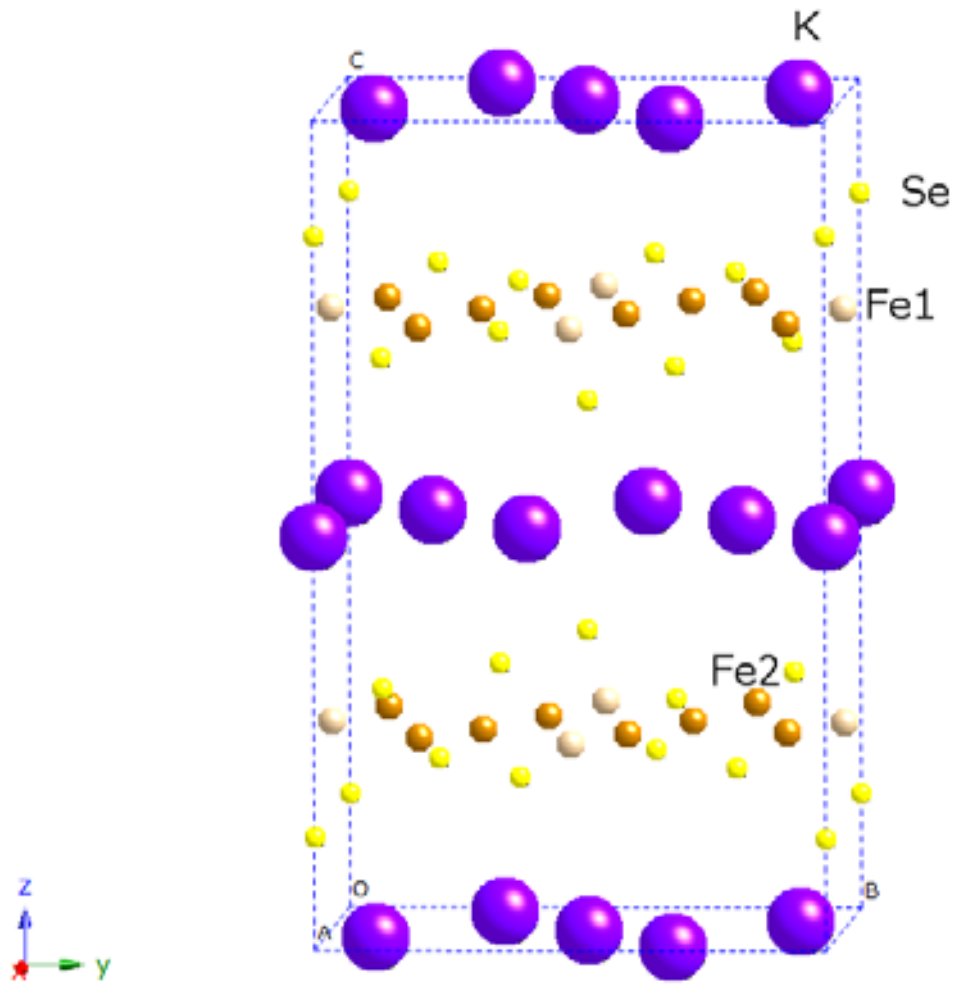

**(b)**

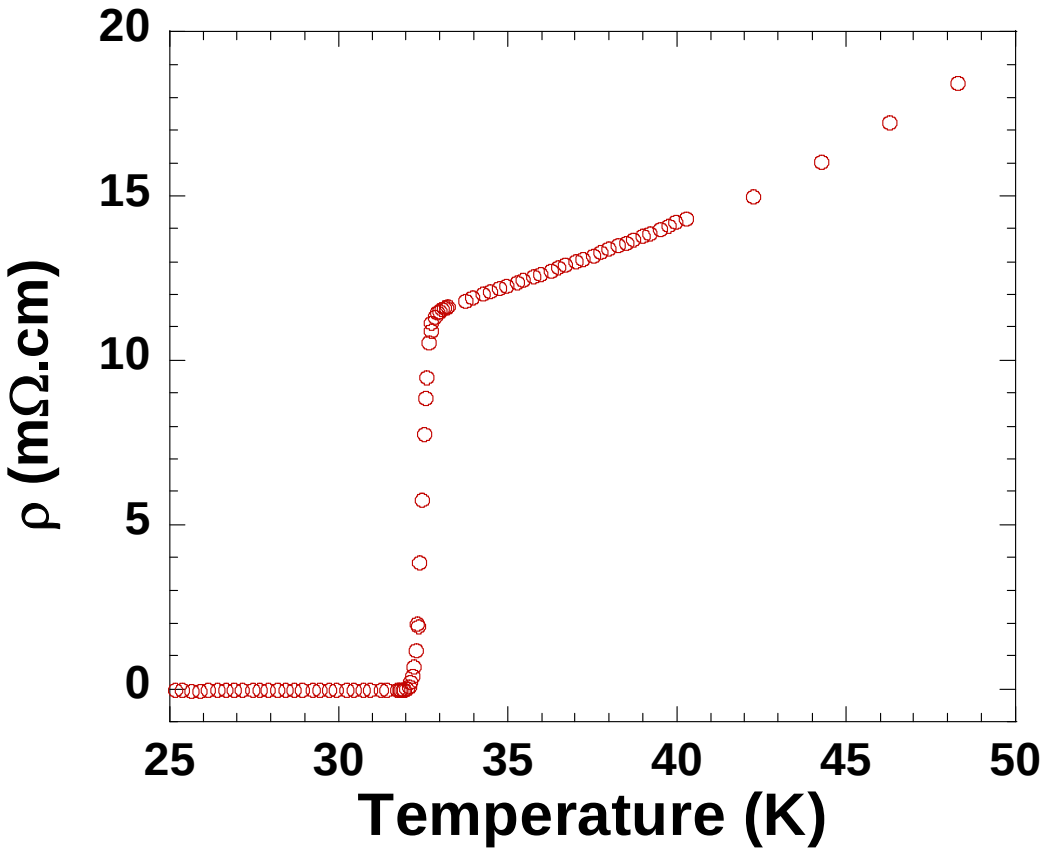

Fig. 2. Tyson *et al.*

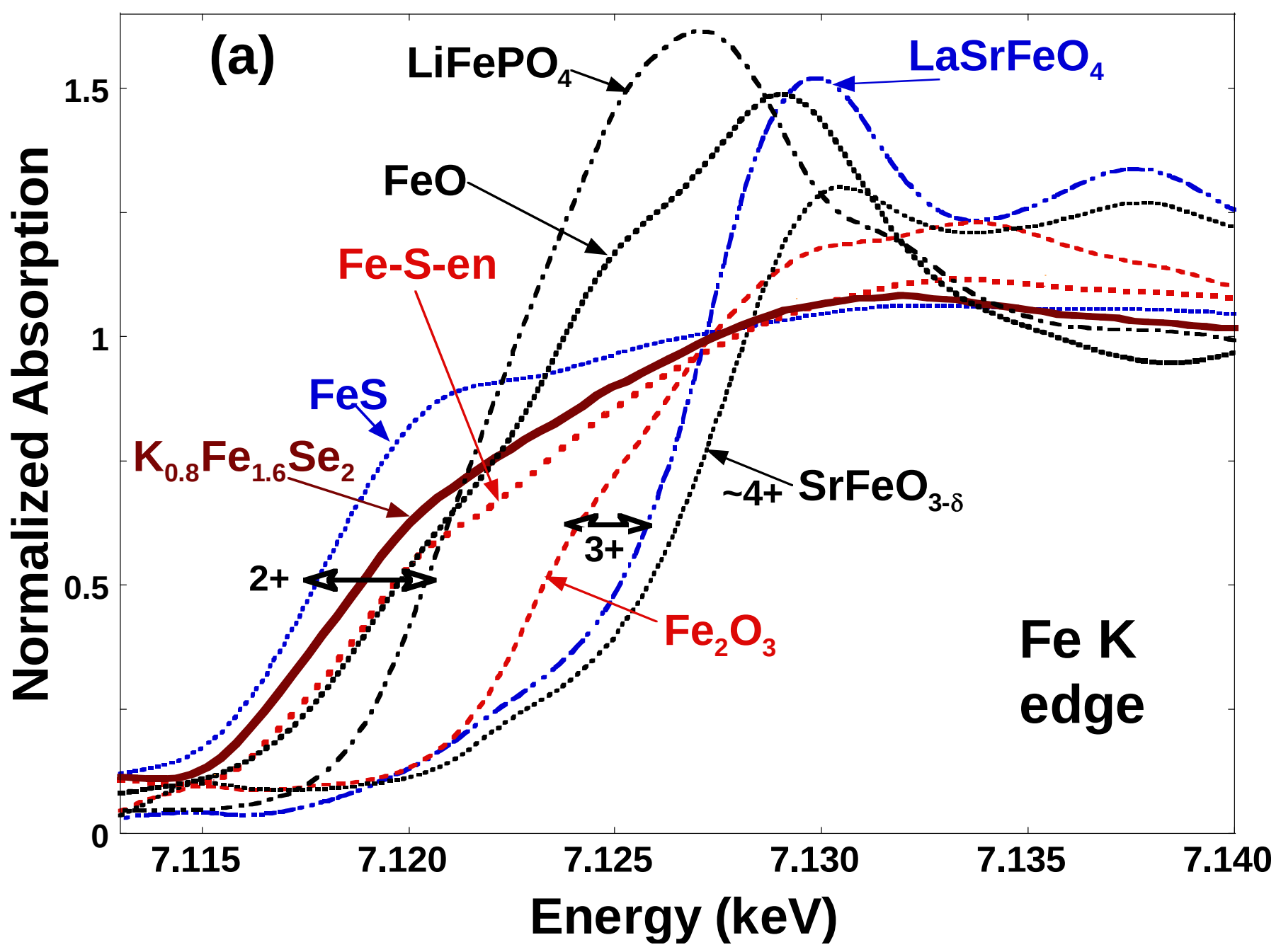

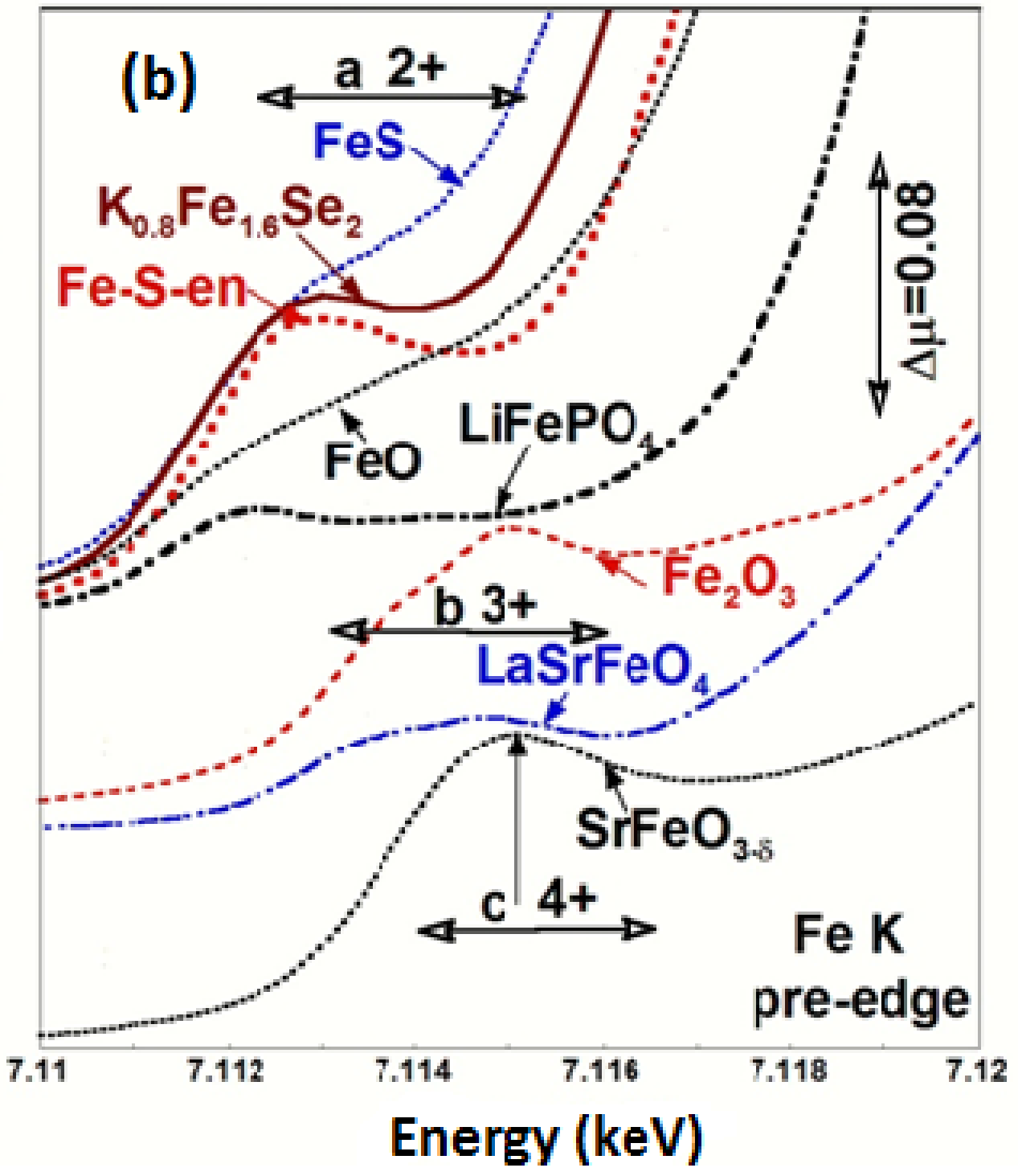

**Fig. 3. Tyson *et al.***

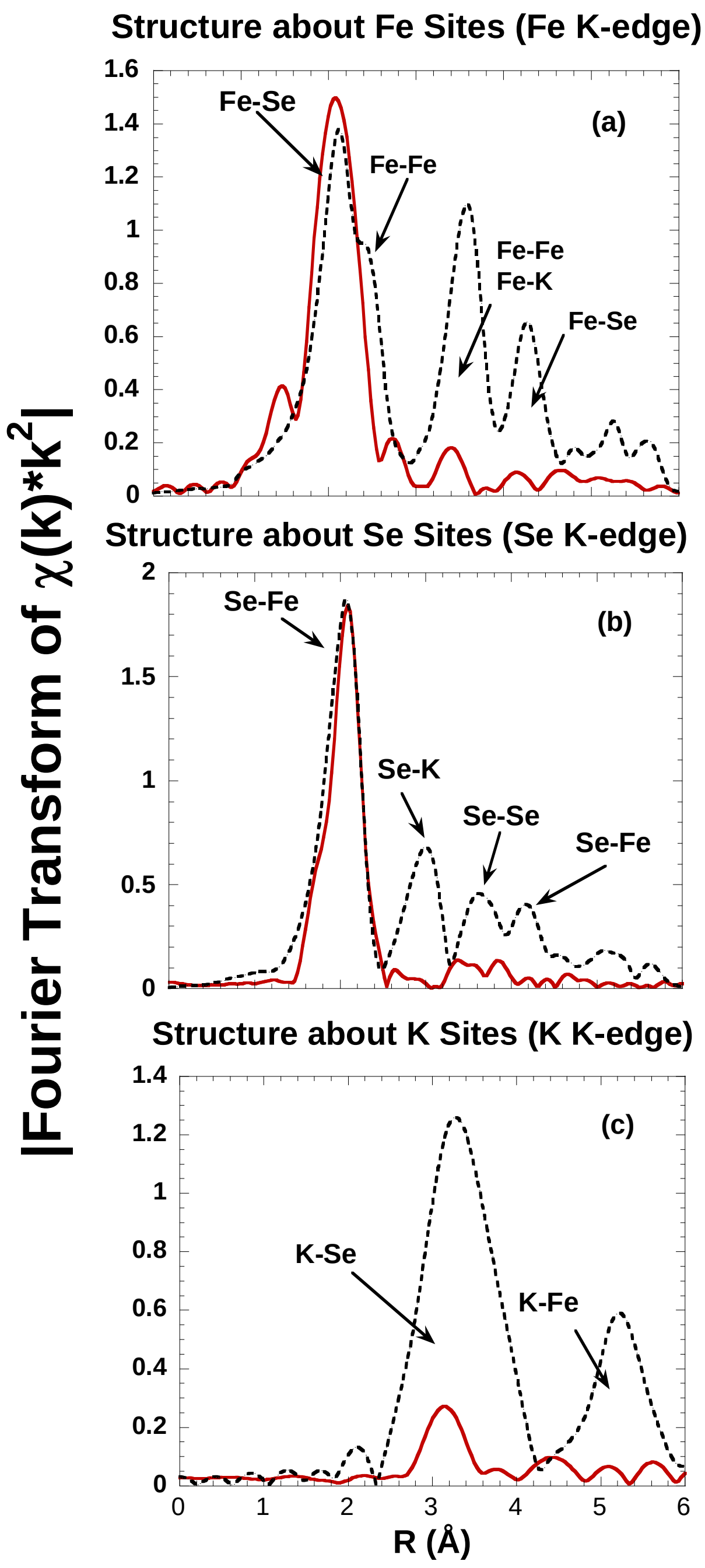

**Fig. 4. Tyson *et al.***

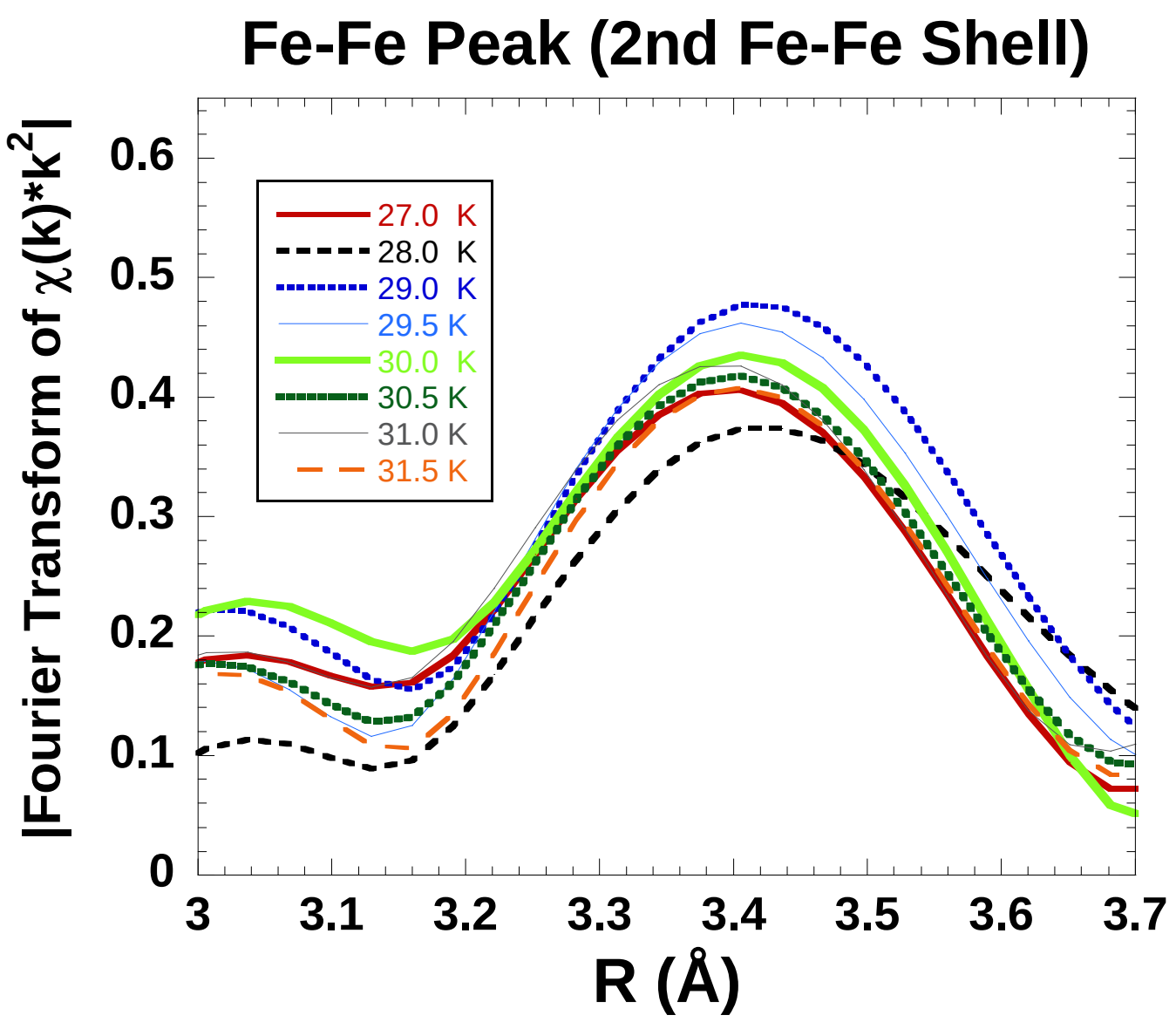

**Fig. 5. Tyson *et al.***

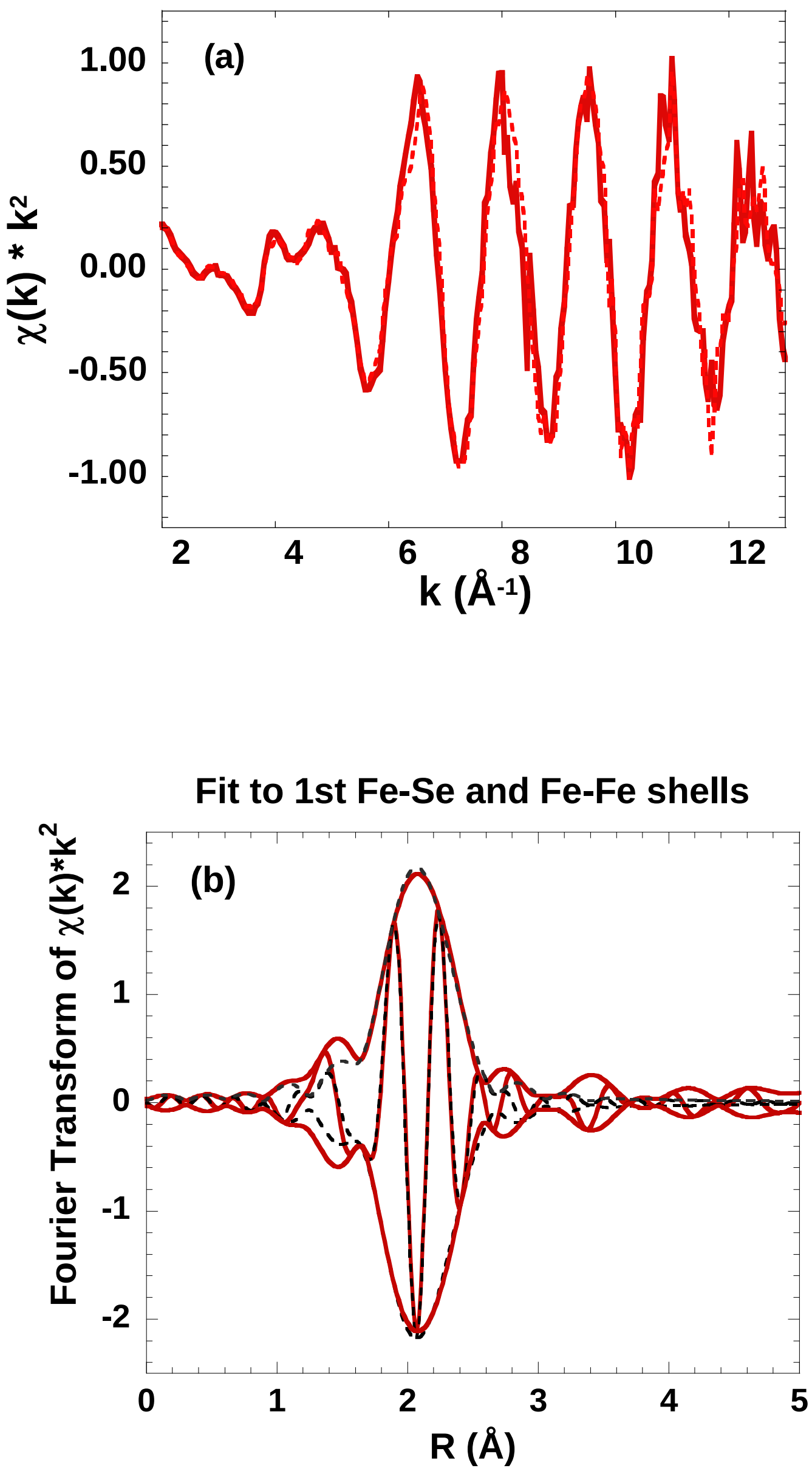

**Fig. 6. Tyson *et al.***

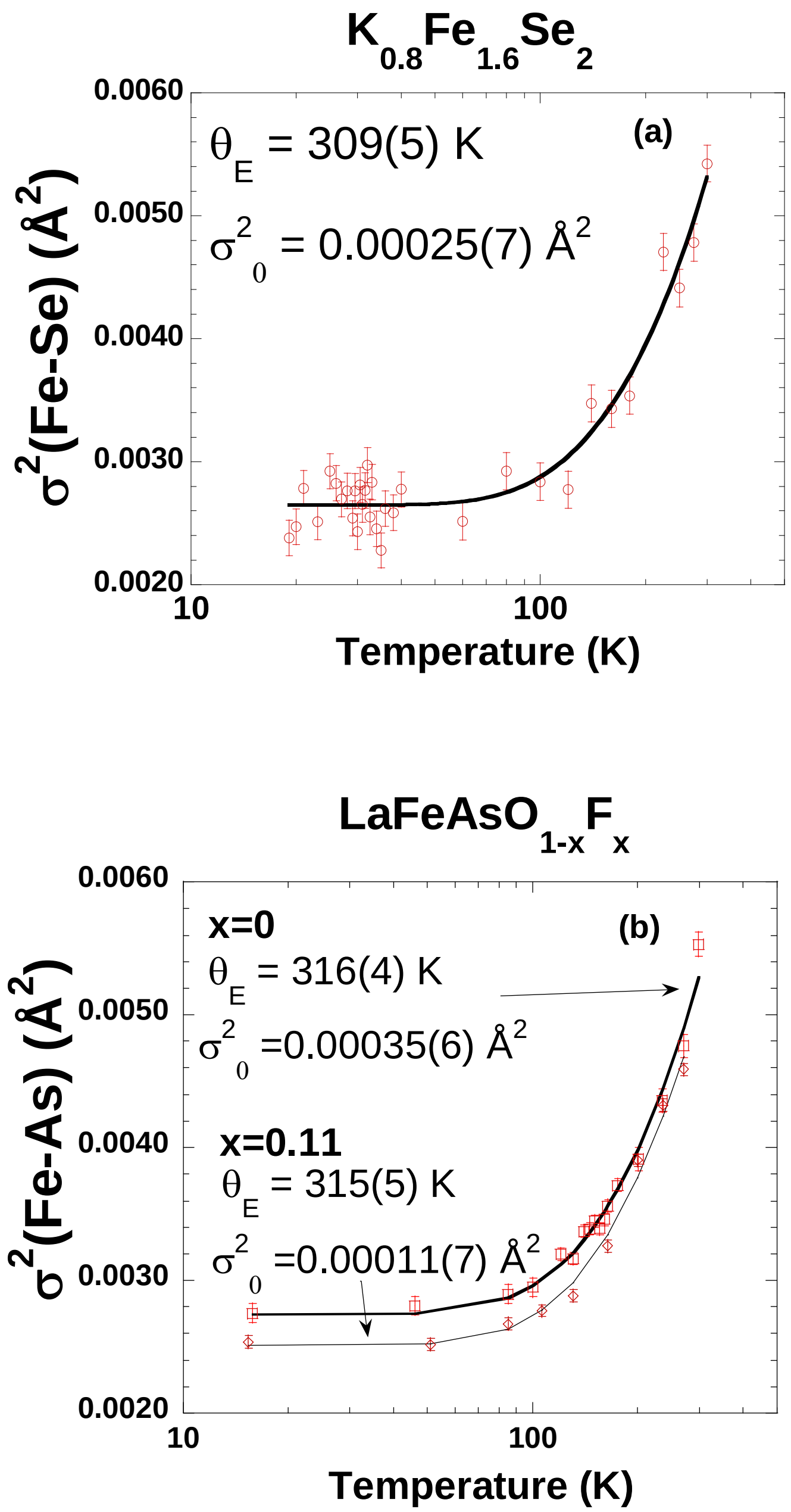

**Fig. 7. Tyson *et al.***

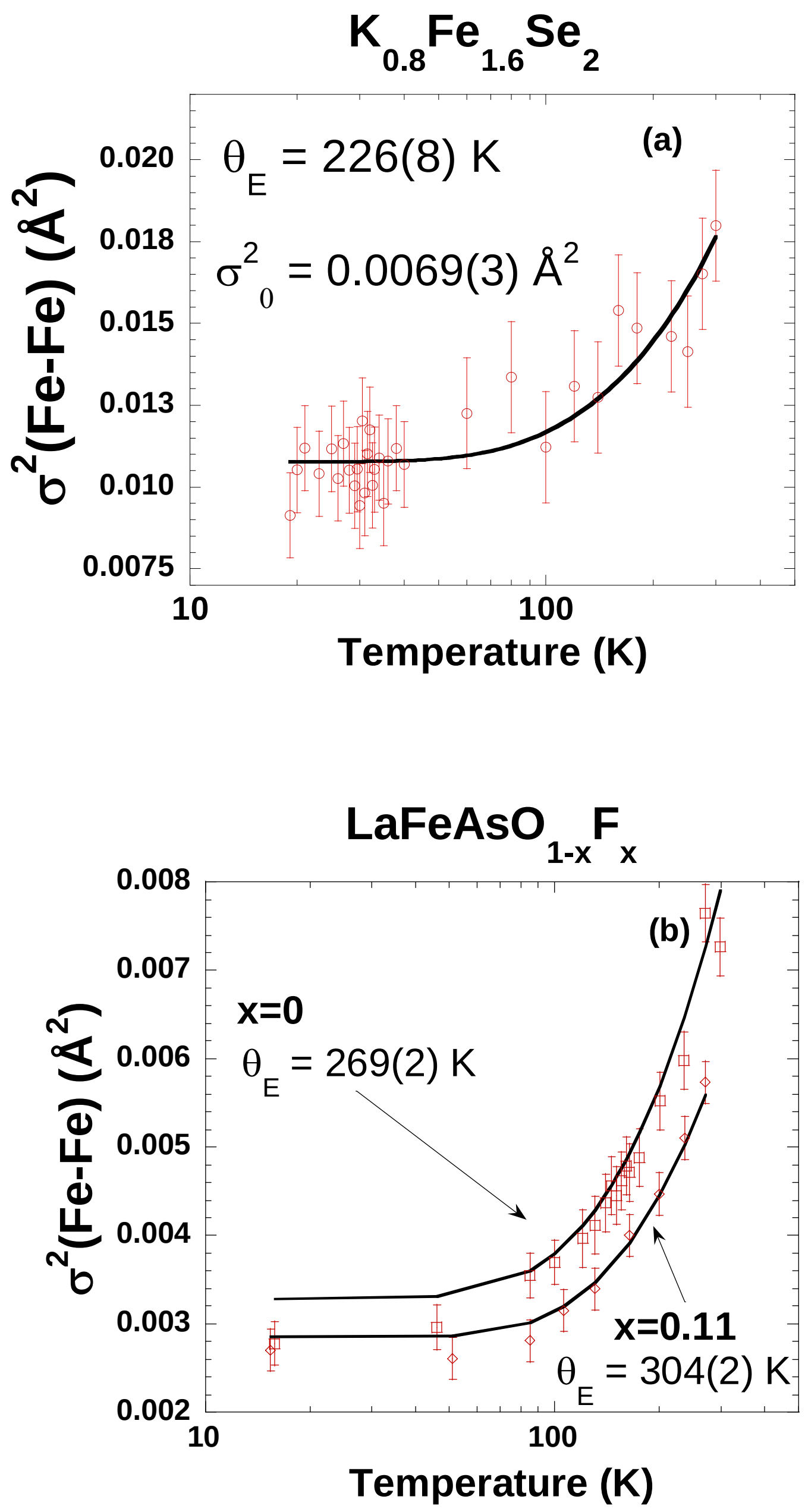

**Fig. 8.  Tyson *et al.***

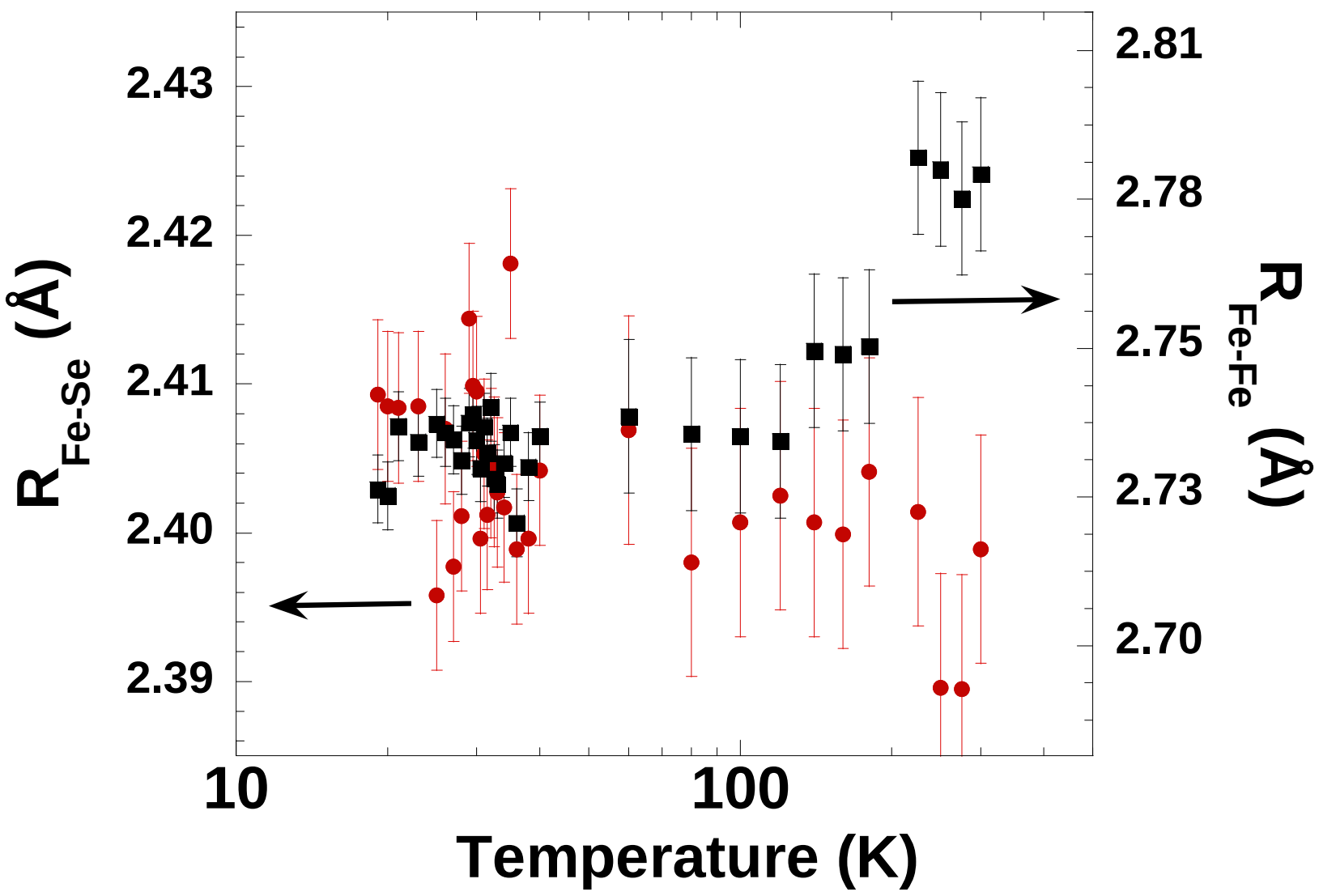